\documentclass[a4paper,11pt]{article}
\pdfoutput=1
\usepackage{graphicx}
\usepackage{amssymb}
\usepackage{amsmath}
\usepackage{subcaption} %
\usepackage{booktabs}
\usepackage{multirow}
\usepackage[title]{appendix}
\usepackage{relsize}
\usepackage{physics} 
\usepackage{enumitem}
\setlist[itemize]{itemsep=0pt,topsep=\baselineskip} %
\usepackage{pgfplots}
\usepackage{pgfplotstable}
\usepackage{tikz,pgfplots}
\usepackage[colorlinks=true, linkcolor=blue, urlcolor=blue, citecolor=blue, linktoc=page]{hyperref}
\usepackage{xcolor}
\pgfplotsset{compat=1.14}
\newsavebox{\foobox}

\newcommand\T{\rule{0pt}{2.3ex}}       %
\newcommand\B{\rule[-1.2ex]{0pt}{0pt}} %

\usepackage{url}
\usepackage{jheppub} %
\newread\imgstream
\immediate\openin\imgstream=imagedata.in
\makeatletter
\def\new@kvginclip#1{}
\def\new@kvgintrim#1{}
\let\old@kvginclip\KV@Gin@clip
\let\old@kvgintrim\KV@Gin@trim
\let\oldincludegraphics\includegraphics
\providecommand{\includegraphics}{}
\renewcommand{\includegraphics}[2][]{%
  \immediate\read\imgstream to \src
  \immediate\read\imgstream to \removecrop
  \ifnum\removecrop=1
      \let\KV@Gin@clip\new@kvginclip
      \let\KV@Gin@trim\new@kvgintrim
  \fi
  \oldincludegraphics[#1]{\src}%
  \let\KV@Gin@clip\old@kvginclip
  \let\KV@Gin@trim\old@kvgintrim}
\makeatother

\begin{document}

\pagenumbering{arabic}
\centerline{\LARGE EUROPEAN ORGANIZATION FOR NUCLEAR RESEARCH}
\vspace{10mm}
\begin{flushright}
CERN-EP-2026-214 \\
July 21, 2026 \\
-- \\
Revised version: \\
\today \\
 \vspace{2mm}
    \vspace{5pt} %
 \vspace{2mm}
\end{flushright}

\begin{center}
\Large{
\bf Measurement of the branching ratio of \break 
the {\boldmath $K^{+}\rightarrow\pi^{+}\nu\bar{\nu}$} decay
\\
\vspace{10mm}
}
\begin{NoHyper}
\renewcommand{\thefootnote}{\fnsymbol{footnote}}
The NA62 Collaboration
\footnote[1]{
Corresponding authors: X.~Chang, R.~Fiorenza, J.~Swallow, \\
email: xiafei.chang@cern.ch, renato.fiorenza@cern.ch, joel.christopher.swallow@cern.ch
}
\end{NoHyper}

\end{center}
\vspace{5mm}

\begin{abstract}

\centerline{\textbf{Abstract}}%

\vspace{5mm}

The NA62 experiment has measured the branching ratio of the rare decay $K^+\to\pi^+\nu\bar{\nu}$ using data collected in 2023 and 2024 at the CERN SPS, obtaining $\mathcal{B}_{\rm 2023-2024}(K^+\to\pi^+\nu\bar{\nu}) = \left(7.2^{+2.3}_{-2.1}\right)\times10^{-11}$.
Including results from data collected between 2016 and 2022, the analysis gives $\mathcal{B}_{\rm 2016-2024}(K^+\to\pi^+\nu\bar{\nu}) = \left(9.6^{+1.9}_{-1.8}\right)\times10^{-11}$, corresponding to a relative precision of 20\%. 
This value is consistent with Standard Model predictions.

\end{abstract}
\vspace{10mm}

\begin{center}
\em{Accepted for publication in JHEP}
\end{center}

\clearpage

\newpage
\clearpage

\hrule
\tableofcontents

\vspace{0.5cm}
\hrule

\section{Introduction}
\label{sec:Intro}

The $K^{+}\rightarrow\pi^{+}\nu\bar{\nu}$ decay is an ultra-rare process and a ``golden mode'' %
in the search for new physics in the quark-flavour sector.
Recent Standard Model (SM) predictions of its branching ratio, $\mathcal{B}(K^{+}\rightarrow\pi^{+}\nu\bar{\nu})$, which differ in the treatment of the CKM parameters, give values of approximately $8\times10^{-11}$ with a precision better than $8\%$~\cite{Buras:2022wpw,KaonsAtCERN2023,Allwicher:2024ncl}, which is currently limited by experimental inputs.
The intrinsic theoretical uncertainty, dominated by the knowledge of long-distance charm contributions, is %
$3\%$~\cite{Brod:2010hi}.
Several models involving physics beyond the SM allow for significant deviations from these predictions, 
correlated with the branching ratios of the $K_L\to\pi^0\nu\bar\nu$ decay and other flavour-changing neutral current processes (e.g. $B$ meson decays)~\cite{Buras:2015yca,Bordone:2017lsy,Bobeth:2017ecx,Aebischer:2020mkv,Deppisch:2020oyx,Descotes-Genon:2020buf,Marzocca:2021miv,Crosas:2022quq,Gorbahn:2023juq,Buras:2024ewl}.
An upper limit of $\mathcal{B}(K_L\to\pi^0\nu\bar\nu)<2.2\times10^{-9}$ at $90\%$ CL has been obtained by the KOTO experiment at J-PARC~\cite{KOTO:2024zbl}.

The NA62 experiment at the CERN SPS was designed and built with the primary objective of measuring $\mathcal{B}(K^{+}\rightarrow\pi^{+}\nu\bar{\nu})$.
Using data collected between 2016 and 2022, NA62 observed $51$ candidate events with an estimated background of $18^{+3}_{-2}$ events, therefore excluding the background-only hypothesis with a significance exceeding $5\sigma$ and establishing the first observation of this decay.
With the 2016--2022 dataset, the branching ratio was measured to be $\left(13.0^{+3.3}_{-3.0}\right)\times10^{-11}$~\cite{Pnn2122paper}.

In the following sections, a measurement of $\mathcal{B}(K^{+}\rightarrow\pi^{+}\nu\bar{\nu})$ using the dataset collected in 2023--2024 is reported, as well as the combination with previous measurements.

\section{Beamline and detector}
\label{sec:Detector}

\begin{figure}[t]
    \centering
    \includegraphics[width=1\linewidth]{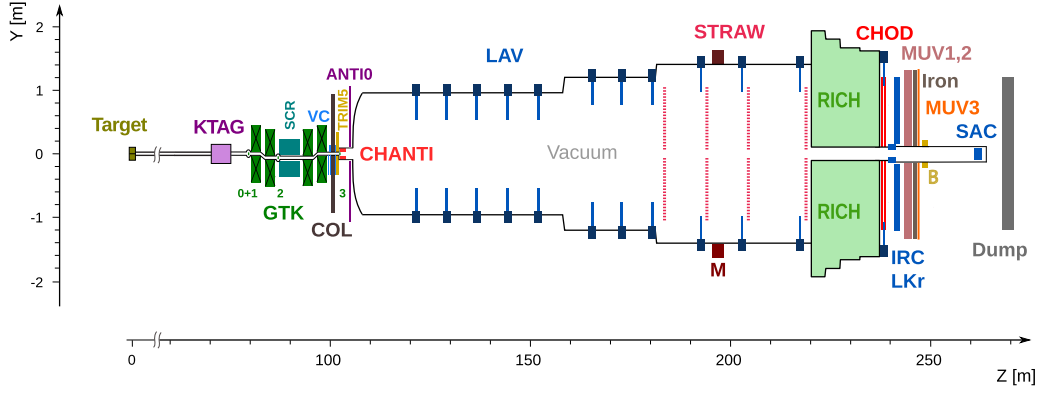}
    \caption{Schematic side view of the NA62 detector for data taking from 2021 onwards. %
    }
    \label{fig:BealineAndDetectorSketch}
\end{figure}

The NA62 beamline and detector are described in detail in~\cite{NA62DetectorPaper,Pnn2122paper}.
The setup used for the collection of this dataset is displayed in figure~\ref{fig:BealineAndDetectorSketch}.
Here, a brief summary is presented, highlighting the improvements introduced in 2023 and 2024.

An unseparated $75$~GeV$/c$ beam of $70\%$ $\pi^{+}$, $23\%$ protons and $6\%$ $K^{+}$ is created by directing $400$~GeV protons extracted from the CERN SPS onto a beryllium target in spills of $4.8$~s duration. The target defines the origin of the right-handed coordinate system shown in figure~\ref{fig:BealineAndDetectorSketch}.

Beam kaons are tagged by a differential Cherenkov counter (KTAG). The KTAG was upgraded before the 2023 data-taking~\cite{CedarH} to use ${\rm H}_{2}$ gas at a pressure of $3.85$~bar as a radiator medium, whereas the previous radiator gas was ${\rm N}_2$ at $1.71$~bar. 
This upgrade lowered the KTAG material budget from $3.9\%$ to $0.7\%$ of a radiation length, leading to a reduction 
in the flux of beam particles scattered into the downstream detector acceptance by 30\%.
The light yield also increased by $14\%$, improving the $K^{+}$ identification efficiency from $99.5\%$ to $99.7\%$ and the time resolution from $71$~ps to $66$~ps.

The three-momenta of beam particles are measured by the Gigatracker (GTK) spectrometer. The veto counter (VC), ANTI0, and CHANTI detectors reject the decays and interactions of beam particles occurring between the first and last GTK stations.

A new prototype FPGA-based TDC system, read out by an ATLAS FELIX board~\cite{FELIX}, became the main readout of the VC in 2023; its use was extended to the CHANTI in 2024. 
This new system allows higher data throughput, avoiding the bandwidth limitations of the previously used TEL62 boards~\cite{Ammendola:2019skv} while providing improved reliability and radiation tolerance.

The beam is delivered into a vacuum tank containing the fiducial volume (FV), defined as the region $105\,\text{m}<Z<170\,\text{m}$. 
Downstream of the FV, 
a magnetic spectrometer (STRAW) measures secondary charged particle momenta, a ring-imaging Cherenkov detector (RICH) provides precise timing and particle identification, and  
scintillator hodoscopes (CHOD) provide trigger signals and time measurements.
Additional particle identification capabilities are provided by a liquid krypton electromagnetic calorimeter (LKr), two sampling hadronic calorimeters (MUV1,2), and an array of scintillator tiles (MUV3) located behind an iron wall.

The photon veto system consists of twelve ring-shaped large-angle veto (LAV) stations, the LKr, the intermediate-ring calorimeter (IRC), and the small-angle calorimeter (SAC). This setup provides hermetic coverage of polar angles in the forward direction between $0$ and $50$~mrad with respect to the beam axis.

\section{Data sample and trigger}
\label{sec:Dataset}

The data sample used for this analysis consists of $8.9\times10^{5}$ spills,
collected in 2023 and 2024.
The expected number of SM signal events per spill (calculated as reported in section~\ref{sec:SignalSensitivity}) is $2.6\times10^{-5}$, $5\%$ larger than for the 2022 dataset. 

Studies of the signal yield per spill as a function of the beam intensity were performed in 2023. The results show a broad maximum around $450$~MHz, occurring due to losses from dead time in the data acquisition and a higher rate of accidental vetoes from both the trigger and offline selections.
Therefore, 
from August 2023, the experiment was operated at $75\%$ of the maximum design beam intensity of about $600$~MHz.
The 2023--2024 dataset has an instantaneous beam intensity, as measured by counting out-of-time GTK signals, of $440$~MHz on average, with a standard deviation of $180$~MHz. For comparison, the 2021--2022 dataset was collected at the maximum design intensity, with an average of $580$~MHz.

A two-stage trigger system is employed with successive hardware (L0) and software (L1) levels~\cite{NA62TriggerPaper}.  
Three trigger lines are used to collect data for this study.
The minimum bias (MB) trigger line, downscaled by a factor of $600$, is used to collect control samples.
The normalisation (NORM) trigger line, downscaled by a factor of $400$, is designed to select a beam particle identified as a kaon and one positively charged STRAW track not identified as a muon; it is used to collect a normalisation sample of $K^+\to\pi^+\pi^0$ decays. The signal-collection (PNN) trigger line, used to collect $K^+\to\pi^+\nu\bar\nu$ signal candidates, is designed to be stricter than the NORM trigger line and places additional veto conditions on extra activity in CHOD, LKr, LAV, and STRAW.

Since 2023, the PNN trigger line has made use of an improved L1 LAV veto condition.
In particular, slewing corrections were implemented in the online reconstruction, allowing the LAV veto windows to be decreased from $6$ to $4$~ns. 
Additionally, the LAV stations upstream of the decay vertex, reconstructed as the mid-point of the closest distance of approach (CDA) segment between the STRAW track and the nominal beam direction, were removed from the coincidence requirement.

\section{Event reconstruction improvements}
\label{sec:RecoUpdates}

New techniques based on machine learning are used for 4D beam particle tracking in the GTK and for particle identification in the calorimetric system, as described below. The new beam particle tracking technique replaces a conventional tracking algorithm, while the particle identification algorithm previously relied on a boosted-decision-tree (BDT) classifier with a smaller set of inputs~\cite{Pnn2122paper}.

\subsection{Beam particle tracking}

The 4D GTK beam particle tracking algorithm is now based on deep learning: %
its architecture consists of a transformer encoder~\cite{TransformerML} and a binary classifier.
In the event reconstruction algorithm, GTK signals from each station are first grouped into space-time clusters.
The model then assigns a score to the connection between each pair of signal clusters in successive GTK stations, and is trained to connect clusters associated to the same particle. 
The binary classifier first discards connections with a score below a threshold.
Then the connections with the highest score are selected, ensuring that tracks do not share clusters.

Training was performed using simulations of the beam
with an intensity profile matching the 2023--2024 dataset.
Signals were not clustered during training, and connections to any signal produced by the same particle were labelled as correct, whereas during reconstruction clustering is performed before tracking: this strategy makes the model more likely to ignore small deviations from the true position of the particle and therefore be less sensitive to clustering errors.
A strong class imbalance arises because the number of pairs of uncorrelated clusters from combinatorics is much larger than the number of pairs of clusters belonging to single tracks. A focal loss function~\cite{FocalLoss1}, down-weighting well-classified pairs, was therefore used for the training.

\begin{figure}[tb]
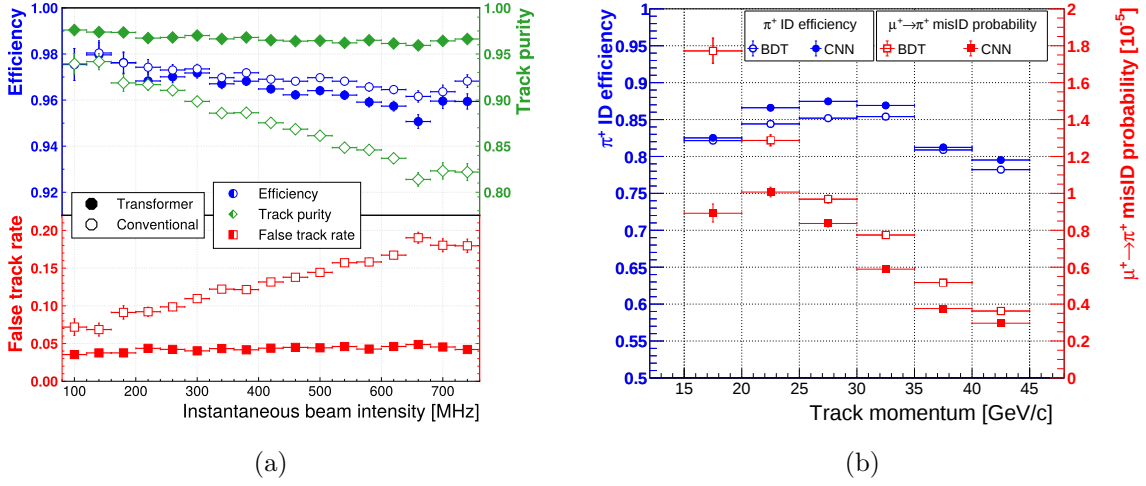

    \centering
    \begin{subfigure}{0.475\textwidth}
        \centering
        \includegraphics[width=\linewidth]{Figures/gtk_recos_v3.fixed.pdf}
        \caption{}%
        \label{fig:GTKreco}
    \end{subfigure}
    \hfill
    \begin{subfigure}{0.49\textwidth}
        \centering
        \includegraphics[width=\linewidth]{Figures/ExtraResultsPlots2324__PIDstudies_PpiPID_Calo_BDTvsCNN_vspbins_FINAL.pdf}
        \caption{}%
        \label{fig:CNNpid}
    \end{subfigure}
    \caption{
    (a) Performance 
    of the new beam particle tracking algorithm (filled markers) compared to the previously used conventional algorithm (empty markers), as a function of instantaneous beam intensity.
    (b) Particle identification performance of the CNN-based classifier (filled markers) compared to the performance of the previously used BDT-based classifier (empty markers), as a function of track momentum. 
    }
    \label{fig:CNN}
\end{figure}

Performance studies using these simulations
show an improvement in the 
false track rate 
(fraction of true tracks which are incorrectly reconstructed)
from $13\%$ to $4\%$,
and in track purity 
(fraction of reconstructed tracks which are correctly reconstructed)
from $87\%$ to $97\%$, 
with a loss of efficiency (fraction of true tracks which are correctly reconstructed) of less than $1\%$ when compared to the previously used conventional tracking algorithm (figure~\ref{fig:GTKreco}). 

\subsection{Calorimetric particle identification}
\label{sec:CNNPID}

The calorimetric particle identification has been improved by implementing a classifier that combines a convolutional neural network (CNN) and a fully connected feed-forward neural network. The inputs to this classifier are based on signals in the LKr and MUV1,2 calorimeters.

Earlier studies showed that the particle identification performance can be improved with the use of a CNN~\cite{NA62_PID_CNN}, relative to that of the original BDT algorithm. 
Building on these studies, a classifier was developed for the present analysis using a ResNet-based~\cite{ResNet} architecture with the Mish activation function~\cite{Mish}.
The input to the CNN is an image with a depth of $10$ channels, representing the energy deposited in the calorimeters divided by the track momentum.
Six channels represent the energy deposits in a $22\times22$ grid of LKr cells around the centre of the energy cluster, in six non-uniform time ranges within $10$~ns of the CHOD time.
The remaining four channels represent the energy deposits in horizontal and vertical slabs of the MUV1 and MUV2 calorimeters, within $10$~ns of the CHOD time.
In the fully connected part of the architecture, the inputs include the fraction of energy deposited in each detector and the size of the LKr cluster.

Training samples of $K^{+}\rightarrow\pi^{+}\pi^{0}$ and $K^{+}\rightarrow\mu^{+}\nu$ decays are selected using the MB trigger line, providing samples of $\pi^{+}$ and $\mu^{+}$ with relative cross-contamination of less than $2\times10^{-8}$ and $4\times10^{-8}$, respectively.
Training 
was performed using the weighted Adam optimiser~\cite{AdamW} with a focal loss~\cite{FocalLoss}.
To improve generalisation and mitigate overfitting, input images were vertically flipped with a probability of $0.5$ during training.

Based on the output of this classifier, a momentum-dependent selection is applied for calorimetric identification of $\pi^{+}$. 
This selection criterion is optimised using the figure of merit $F_{i} = -{A_{\pi\nu\bar{\nu},i}}/({5/2 + \sqrt{B_{i}}})$~\cite{Punzi:2003bu}, where $A_{\pi\nu\bar{\nu},i}$ and $B_{i}$ are the signal acceptance and expected background in bins of momentum, $i$. %

The use of the new classifier increases the pion identification probability by about 
$2\%$, while reducing the $\mu^{+}$ as $\pi^{+}$ misidentification probability by $40\%$ relative, when compared to the previously used BDT classifier
(figure~\ref{fig:CNNpid}).

\section{Event selection}
\label{sec:Selection}

The signal selection for $K^{+}\rightarrow\pi^{+}\nu\bar{\nu}$ decays and the normalisation selection for $K^{+}\rightarrow\pi^{+}\pi^{0}$ decays are applied to data collected with the PNN and NORM trigger lines, respectively.

Both selections require a single, positively charged particle downstream of the FV, identified as a pion by both the RICH and the calorimetric system (according to the algorithm described in section~\ref{sec:CNNPID}). 
A $K^+$ candidate is tagged in the KTAG, and the candidate $\pi^+$ track must match the $K^+$ both in space (using the CDA between the $K^+$ and the $\pi^+$ tracks) and in time (using timing from the GTK, KTAG and RICH).
An additional criterion on the $K$/$\pi$ association uses a Bayesian discriminant which also takes into account the GTK occupancy in the event.
The $K$/$\pi$ matching efficiency, measured in a control sample of $K^+\to\pi^+\pi^+\pi^-$ events, is $79\%$, consistent with the result from the previous dataset~\cite{Pnn2122paper}, while the probability of matching the $\pi^+$ to an accidental GTK track is reduced from $6\%$ to $4\%$.
The reconstructed vertex between the STRAW and the GTK tracks is required to be inside the FV.
Veto conditions based on GTK, VC, CHANTI, and ANTI0 signals, and criteria based on spatial information of the tracks, reject decays
and interactions upstream of the FV.
Photon and multiplicity veto criteria, which reject events with extra activity from additional photons or charged particles, are applied to the signal sample only.
The selection criteria are described further in~\cite{Pnn2122paper}.

The relaxed L1 LAV veto condition (section~\ref{sec:Dataset})
results in a degraded rejection of inelastic interactions of beam particles in GTK3. This rejection is restored and improved by requiring, both in the signal and in the normalization selections, that no signals in LAV1 (the most upstream LAV station) be present within $2$~ns of the $K^{+}$ time.

\begin{figure}[t]
    \centering
        \includegraphics[width=0.7\linewidth]{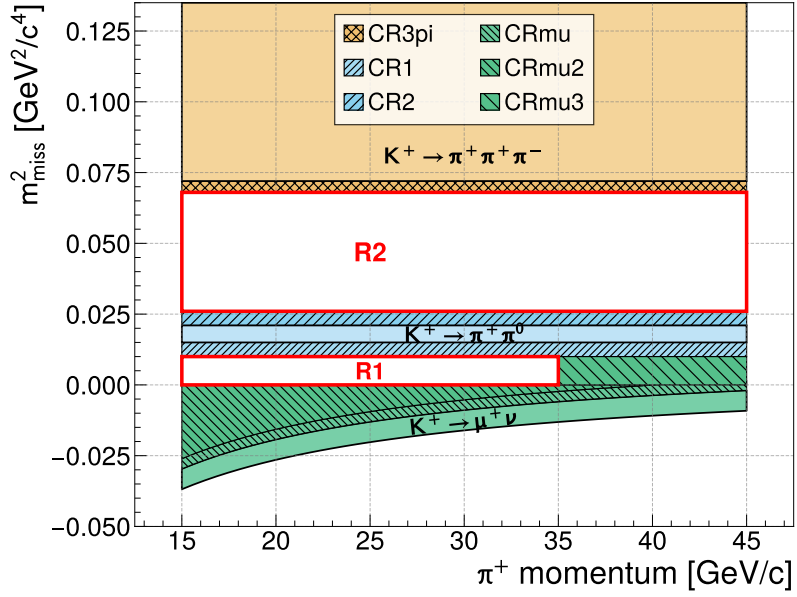}
    \caption{
Definition of signal regions (R1 and R2), background regions (labelled by the corresponding decay modes), and control regions (CR3pi, CR1, CR2, CRmu, CRmu2, CRmu3) in the $(p_{\pi},m_{\rm miss}^{2})$ plane. CR3D overlaps R1 and R2 in this projection.
    }
    \label{fig:kine-regions}
\end{figure}

The squared missing mass is calculated as $m_\text{miss}^2=(P_K-P_\pi)^2$, where $P_K$ and $P_\pi$ are the 4-momenta of the $K^+$ and $\pi^+$ candidates, respectively. Normalisation $K^{+}\rightarrow\pi^{+}\pi^{0}$ events are required to have $m_\text{miss}^2$ compatible with $m_{\pi^0}^2$~\cite{PDG}. %
Kinematic regions are defined in terms of $m_\text{miss}^2$ and the candidate $\pi^+$ momentum $p_\pi$: two \emph{signal regions} (R1 and R2), where events passing the signal selection are kept masked until completion of the analysis; \emph{background regions}, 
containing $K^+\to\mu^+\nu$, $K^+\to\pi^+\pi^0$ and $K^+\to\pi^+\pi^+\pi^-$ decays, respectively; and \emph{control regions}, used to validate background estimates. They are shown in figure~\ref{fig:kine-regions}.
The definition of the signal regions includes additional constraints based on squared missing mass observables calculated analogously to $m_{\rm miss}^{2}$ after replacing the $K^{+}$ momentum with the average beam momentum, or  
the $\pi^{+}$ velocity with the velocity measured by the RICH.
An additional control region, CR3D, is defined by the same $(p_{\pi},m_{\rm miss}^{2})$ conditions as the signal regions but inverting the additional conditions on these alternative squared missing mass observables.

The $\pi^{0}$ rejection inefficiency, defined as the probability that a $K^{+}\rightarrow\pi^{+}\pi^{0}$ candidate passing the normalisation selection is not rejected by the photon or multiplicity veto criteria, is measured 
to be $(1.24\pm0.04)\times10^{-8}$. This is $30\%$ better than for 2021--2022 data due to improved data quality control and optimised beam intensity.

\section{Signal sensitivity}
\label{sec:SignalSensitivity}

The effective number of $K^+$ decays is evaluated as
\begin{equation}
    N_{K} = \frac{N_{\pi\pi}\, D_\text{NORM}}{\mathcal{B}_{\pi\pi}\,A_{\pi\pi}}\,,
\end{equation}
where $N_{\pi\pi}$ is the number of normalisation $K^+\to\pi^+\pi^0$ events selected, 
$D_\text{NORM}=400$ is the downscaling of the NORM trigger line,
$\mathcal{B}_{\pi\pi}=(20.67\pm0.08)\%$ is the branching ratio of the $K^+\to\pi^+\pi^0$ decay~\cite{PDG}, and $A_{\pi\pi}$ is the normalisation selection acceptance. The relative background contamination of the normalisation sample from other kaon decays, estimated with simulations, is less than $10^{-4}$.

The single event sensitivity (SES) is defined as 
\begin{equation}
    \mathcal{B}_\text{SES}(p_{i}) =
   \frac{\mathcal{B}_{\pi\pi} A_{\pi\pi}(p_{i})}{N_{\pi\pi}(p_{i}) \, D_{\text{NORM}}}
   \frac{1}{
    A_{\pi\nu\bar{\nu}}(p_{i})\,
    \varepsilon_\text{trig}(p_{i})\,
    \varepsilon_\text{RV}}\,\,,
\end{equation}
where $p_{i}$ indicates one of the six $5$~GeV/$c$ wide bins of $\pi^+$ momentum in the range 15--45~GeV/$c$,
$A_{\pi\nu\bar\nu}$ is the signal selection acceptance, 
$\varepsilon_\text{trig}$ is the ratio of trigger efficiencies for the PNN and NORM trigger lines for the signal and normalisation selections respectively, 
and $1-\varepsilon_\text{RV}$ is the probability that a signal event is rejected by the photon and multiplicity veto conditions due to the presence of accidental activity.
The latter is independent of $p_\pi$ and only depends on the beam intensity. These quantities and their uncertainties are measured as described in~\cite{Pnn2122paper}.
The dominant uncertainty is systematic, accounting for effects in the photon and multiplicity veto conditions which do not cancel in the acceptance ratio, and amounts to $4\%$.
The expected number of SM $K^{+}\rightarrow\pi^{+}\nu\bar{\nu}$ events is calculated as
\begin{equation}
    N_{\pi\nu\bar{\nu}}^{\text{SM}}(p_{i}) = \frac{\mathcal{B}_{\pi\nu\bar{\nu}}^{\text{SM}}}{\mathcal{B}_\text{SES}(p_{i})}\,\,.
    \label{eqn:Npnn}
\end{equation}
The benchmark value $\mathcal{B}^\text{SM}_{\pi\nu\bar\nu}=8.4\times10^{-11}$~\cite{Buras:2015qea} is used as in previous NA62 papers.

Table~\ref{tab:SESSummaryTable2324} reports each quantity, summed or averaged over momentum bins. 
The lower beam intensity 
in the 2023--2024 dataset leads to an improvement in $\varepsilon_\text{RV}$ by a factor of $1.14$ with respect to the 2021--2022 dataset~\cite{Pnn2122paper}.

\begin{table}[htb] 
    \centering
    \caption{
    Inputs to the signal sensitivity for the 2023--2024 data sample. 
    The number of expected SM $K^+\to\pi^+\nu\bar\nu$ events is evaluated by equation~\eqref{eqn:Npnn} with $\mathcal{B}_{\pi\nu\bar{\nu}}^{\text{SM}} = 8.4\times10^{-11}$.}
        \vspace{8pt}
    \begin{tabular}{|cl|c|} 
    \hline
     & Quantity & Value \T\B \\
    \hline
    \hline 
    $N_{\pi\pi}$ & Number of normalisation events & $3.927\times10^{8}$ \T\B \\ 
    $A_{\pi\pi}$ & Normalisation acceptance & $(12.971\pm0.009)\%$ \T\B \\
    $N_{K}$ & Effective number of $K^{+}$ decays & $(5.93\pm0.02)\times10^{12}$ \T\B \\ 
    \hline
    $A_{\pi\nu\bar\nu}$ & Signal acceptance & $(7.36\pm0.33)\%$ \T\B \\
    $\varepsilon_\text{trig}$ & Trigger efficiency ratio & $(86.4\pm1.2)\%$ \T\B \\
    $\varepsilon_\text{RV}$ & Random veto efficiency & $(72.3\pm0.7)\%$ \T\B \\ 
    \hline
    $\mathcal{B}_\text{SES}$ & Single event sensitivity & $(3.67\pm0.18)\times10^{-12}$ \T\B \\  
    $N_{\pi\nu\bar{\nu}}^{\text{SM}}$ & Number of expected SM $K^{+}\rightarrow\pi^{+}\nu\bar{\nu}$ events & $22.9\pm1.1$ \T\B \\
    \hline  
    \end{tabular} 
    \label{tab:SESSummaryTable2324}
\end{table}

\section{Backgrounds}
\label{sec:BackgroundStudies}

\def\generalregion{R} %

Background arises either from misreconstructed $K^{+}$ decays in the FV 
or from decays or interactions upstream of the FV. %
The methods used for background estimation, focusing on improvements over the analysis of the previous dataset, are described below.
A summary of the background estimates is given in table~\ref{tab:backgrounds}. 

\subsection{Kaon decays in the fiducial volume}

For each of the main $K^{+}$ decay modes, $K^{+}\rightarrow\mu^{+}\nu$, $K^{+}\rightarrow\pi^{+}\pi^{0}$, and $K^{+}\rightarrow\pi^{+}\pi^{+}\pi^{-}$, a kinematic tail fraction is defined as the ratio between the numbers of events in the signal and background regions, evaluated in a dedicated control sample.
The control samples are defined in such a way that the $m_\text{miss}^2$ distribution is kept unbiased: for $K^{+}\rightarrow\mu^{+}\nu$ and $K^{+}\rightarrow\pi^{+}\pi^{0}$, these samples are selected by inverting PID and photon veto conditions respectively, while for $K^{+}\rightarrow\pi^{+}\pi^{+}\pi^{-}$ the sample is derived from simulations~\cite{Pnn2122paper,Pnn2017paper}.
The values of the kinematic tail fractions are reported in table~\ref{tab:kinematic-tails}. 
Systematic uncertainties of $15\%$ and $30\%$, derived by comparing measurements of the control samples in data and simulations, are included in these results for the $K^{+}\rightarrow\pi^{+}\pi^{0}$ and $K^{+}\rightarrow\mu^{+}\nu$ cases, respectively.
The number of background events is estimated by multiplying the kinematic tail fraction by the number of events satisfying the signal selection in the corresponding background region.

Additional contributions from the radiative decays $K^+\to\pi^+\pi^0\gamma$ and $K^+\to\mu^+\nu\gamma$ are evaluated to be $0.13\pm0.07$ and $0.22\pm0.18$, respectively, using the procedure described in~\cite{Pnn2122paper}.
The contribution from the $K^+\to\mu^+\nu\gamma$ background (arising from 
the reconstruction of the muon and photon in the LKr as a single $\pi^+$-like
cluster),
relative to the expected signal, is approximately $10$ times lower than in the analysis of 2021--2022 data. This is due to the lower occupancy in the MUV1,2 calorimeters at lower beam intensity, as well as to the 
more than $3$
times better rejection power of this background from the new calorimetric identification algorithm (section~\ref{sec:CNNPID}). %

\begin{figure}[htb]
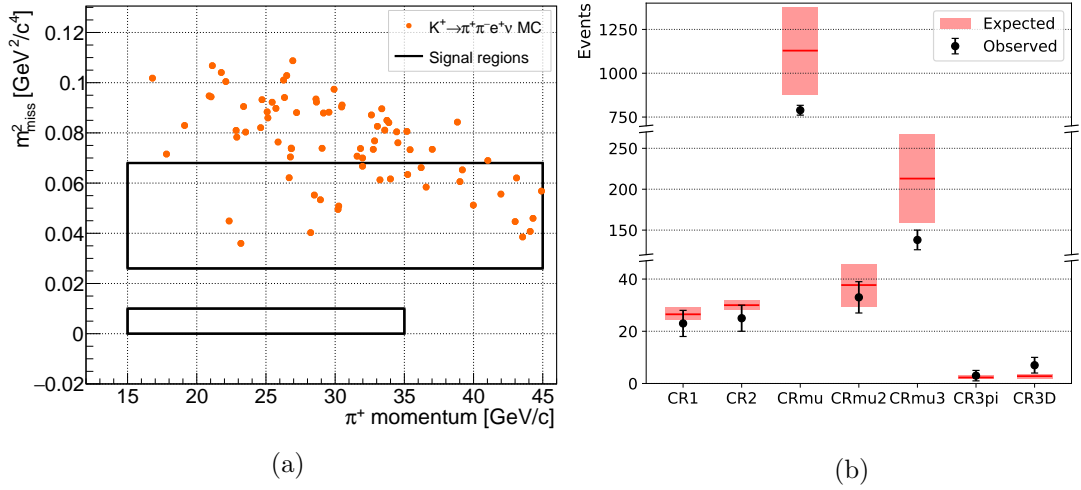

    \centering
    \begin{subfigure}[c]{0.49\textwidth}
        \centering
        \vspace{18pt}
        \includegraphics[width=\linewidth]{Figures/ExtraResultsPlots2324_Ke4MC_mm2_vs_p__SRs_leg.pdf}
        \caption{}%
        \label{fig:Ke4MCevents}
    \end{subfigure}
    \begin{subfigure}[c]{0.48\textwidth}
        \centering
        \includegraphics[width=\linewidth]{Figures/control_regions.pdf}
        \caption{}%
        \label{fig:ControlRegions}
    \end{subfigure}
    \caption{ 
    (a) Distribution in the $(p_{\pi},m_{\rm miss}^{2})$ plane of the events selected from the simulated $K^+\to\pi^+\pi^-e^+\nu$ sample. Each simulated event corresponds to $0.065$ %
    events in the 2023--2024 dataset. 
    (b) Expected and observed numbers of events in the kinematic control regions shown in figure~\ref{fig:kine-regions}, used to validate the backgrounds from $K^+$ decays in the FV. 
        The global $p$-value for the comparison is $0.65$. }
\end{figure}

Simulations are used to evaluate backgrounds from $K^+\to\pi^+\pi^-e^+\nu$, $K^+\to\pi^+\gamma\gamma$ and $K^+\to\pi^0\ell^+\nu$ decays in the FV~\cite{Pnn2122paper}. The largest background from $K^+$ decays in the FV is $K^+\to\pi^+\pi^-e^+\nu$: its contribution is evaluated with a sample of $2.5\times10^9$ simulated events. The distribution of the selected $K^+\to\pi^+\pi^-e^+\nu$ simulated events in the $(p_{\pi^+}, m_\text{miss}^2)$ plane is shown in figure~\ref{fig:Ke4MCevents}.

Background estimates for the main $K^+$ decays in the FV are validated using the kinematic control regions: the results are shown in figure~\ref{fig:ControlRegions}.

\begin{table}[htb] 
    \centering
    \caption{(a) Background estimates for 2023--2024 data, summed over the six $\pi^{+}$ momentum bins. (b) Kinematic tail fractions, averaged over momentum bins.%
    }
    \vspace{8pt}
    \subfloat[]{
    \label{tab:backgrounds}
    \begin{tabular}{|l|c|} 
    \hline
    Background & Events \T\B \\
    \hline
    \hline 
    $K^{+}\rightarrow\pi^{+}\pi^{0}(\gamma)$ & $1.19\pm0.10$  \T\B \\ 
    $K^{+}\rightarrow\mu^{+}\nu(\gamma)$ & $1.39\pm0.29$  \T\B \\ 
    $K^{+}\rightarrow\pi^{+}\pi^{+}\pi^{-}$ & $0.25\pm0.05$ \T\B \\
    $K^{+}\rightarrow\pi^{+}\pi^{-}e^{+}\nu$ & $1.59^{+0.51}_{-0.43}$ \T\B \\
    $K^{+}\rightarrow\pi^{+}\gamma\gamma$ & $0.04\pm0.04$ \T\B \\
    $K^{+}\rightarrow\pi^{0}\ell^{+}\nu$ & $<0.001$  \T\B \\ 
     Upstream & ~$7.5^{+2.9}_{-2.2}$ \T\B \\ %
    \hline
    Total & \hspace{-2pt}$12.0^{+2.9}_{-2.2}$ \T\B \\ %
    \hline
    \end{tabular}
    }
    \hfill
    \subfloat[]{
    \label{tab:kinematic-tails}
    \begin{tabular}{|l|c|c|}
\hline 
Background & Kinematic tail fraction %
\tabularnewline
\hline 
\hline
$K^{+}\to\pi^{+}\pi^{0}$ & $(9.4\pm1.5)\times10^{-4}$ %
\tabularnewline
$K^{+}\to\mu^{+}\nu$ & $(1.4\pm0.5)\times10^{-5}$ %
\tabularnewline
$K^{+}\to\pi^{+}\pi^{+}\pi^{-}$ & $(6\pm1)\times10^{-6}$ %
\tabularnewline
\hline 
\end{tabular}
}
    \label{tab:BackgroundsSummaryTable2324}
\end{table}

\subsection{Upstream background}

\begin{figure}[tb]
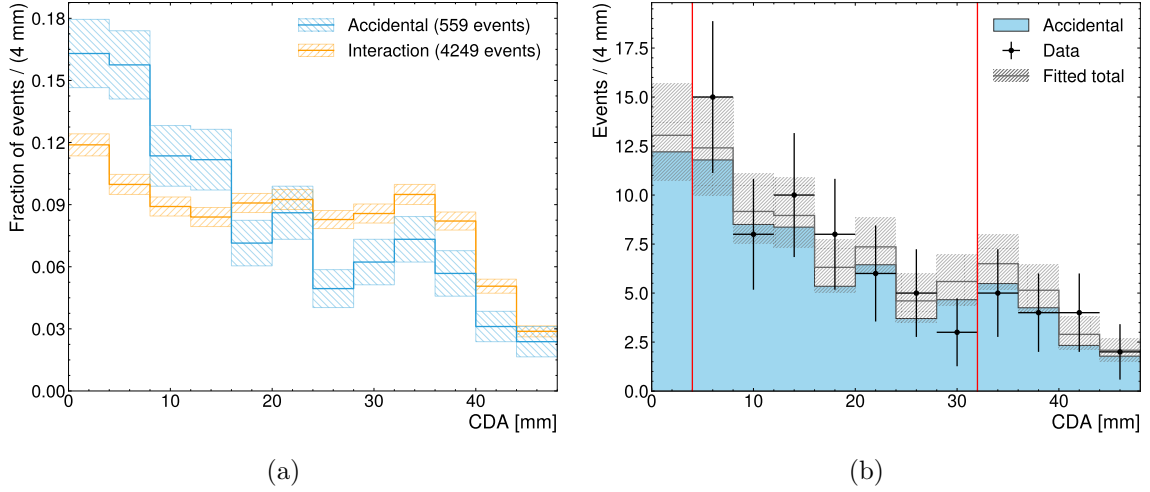

    \centering
    \begin{subfigure}{0.49\textwidth}
        \centering
        \includegraphics[width=\linewidth]{Figures/cda_templates.pdf}
        \caption{}%
        \label{fig:UpsBkgEval_cda_templates}
    \end{subfigure}
    \hfill
    \begin{subfigure}{0.49\textwidth}
        \centering
        \includegraphics[width=\linewidth]{Figures/EvaluateUpstreamBackground_cdaTemplateFit_bkg.pdf}
        \caption{}%
        \label{fig:UpsBkgEval_cda_fit}
    \end{subfigure}
    \caption{
        (a) Normalised CDA distributions of the interaction and accidental samples.
        (b) Result of the template fit to the upstream reference sample. %
        Vertical lines delimit the fit region. The fitted accidental component and total, with its statistical uncertainty, are shown.
    }
    \label{fig:UpsBkgEval}
\end{figure}

The upstream background arises from decays or interactions upstream of the FV by 
two classes of mechanisms. One involves the inelastic interaction of beam particles with material along the beamline. The other involves the decay of a beam $K^+$ upstream of the FV and the matching of a downstream $\pi^+$ to an accidental beam particle. In both cases, a fake vertex can be reconstructed in the FV. 
By modifying the selection criteria, two data samples are obtained that enhance each of the two mechanisms; they are statistically independent of each other and of the signal sample. The modifications for the selection of the \emph{interaction} sample include inverting veto conditions based on CHANTI, LAV1, GTK, ANTI0, and track-spatial information, while for the \emph{accidental} sample they include inverting veto conditions based on GTK, VC, and timing.
The CDA distributions for each of these two samples, which are assumed to be characteristic of the underlying mechanisms, are shown in figure~\ref{fig:UpsBkgEval_cda_templates}.

A fully data-driven strategy is used to evaluate the upstream background. 
An upstream reference sample (URS) of $70$ events is selected by applying the full signal selection, except for the $K$/$\pi$ matching criteria, and inverting the CDA requirement to $\text{CDA}>4$~$\text{mm}$.
A template fit is performed to the CDA distribution of the URS over the range $4$--$32$~$\text{mm}$, where the templates are given by the CDA distributions of the interaction and accidental samples.
The result of the fit (figure~\ref{fig:UpsBkgEval_cda_fit}) is used to compute the factor $f_\text{CDA}$ for the extrapolation of the CDA distribution of the URS to the signal region ($\text{CDA}<4\,\text{mm}$).
A $15\%$ relative statistical uncertainty on $f_{\rm CDA}$ from the fit, consisting of contributions from the templates and the URS distribution, is found using a parametric bootstrapping method: pseudo-experiments are generated from the fitted parametric model, and the fit is repeated to obtain the distributions of the fitted parameters.
An additional systematic uncertainty of $15\%$ is assigned to account for the choice of the data samples which determine the templates, and is obtained by varying their selection conditions. 
The result is $f_\text{CDA} = 0.19^{+0.05}_{-0.04}$. %
This procedure highlights that the accidental component is dominant in the estimated background, as seen in figure~\ref{fig:UpsBkgEval_cda_fit}. 

The probability for an upstream background event to pass the additional Bayesian $K$/$\pi$ matching criterion is evaluated in bins of the relevant variables (related to GTK occupancy and to the relative timing between the $K^+$ and the $\pi^+$). 
This probability, denoted $P^\text{match}_i$ for the $i$-th bin, is extracted from the normalisation sample~\cite{Pnn2122paper} and is found to be $0.68$ on average.
An additional factor $(1+\delta_\text{match})$ accounts for possible bias in the distributions of the $K$/$\pi$ matching variables in the URS with respect to the background. The correction term $\delta_\text{match}=-0.13\pm0.13_\text{syst}$ is evaluated from the fraction of events passing the $K$/$\pi$ matching criterion in the interaction and accidental samples satisfying $\text{CDA}<4\,\text{mm}$, with the relative weights of the interaction and accidental components of the URS applied.
Finally, the upstream background is evaluated as
\begin{equation}
    N_\text{ups} = (1+\delta_\text{match}) f_\text{CDA}\sum_{i} N^\text{URS}_i P^\text{match}_i = 7.5^{+2.9}_{-2.2} \,\,,
\end{equation}
where $N^\text{URS}_i$ is the number of URS events in the $i$-th bin.

\begin{figure}[t]
    \centering
        \centering
        \includegraphics[width=0.5\linewidth]{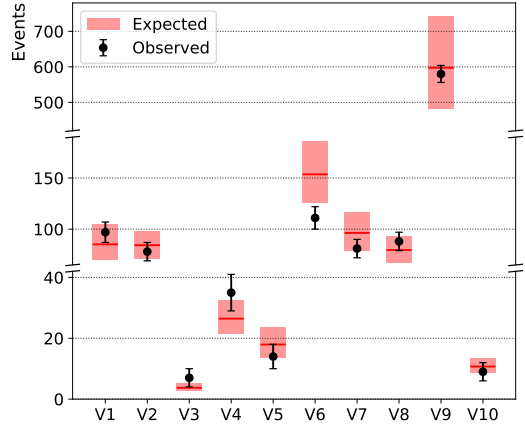}
     
    \caption{
         Expected and observed numbers of events in the samples used to validate the upstream background estimate.
        The global $p$-value for the comparison is $0.79$.
    }
       \label{fig:ValidationRegions}
\end{figure}

The upstream background estimate is validated with ten statistically independent data samples, each enhancing particular upstream background mechanisms~\cite{Pnn2122paper}. The results of the validation are shown in figure~\ref{fig:ValidationRegions}. 
The good agreement between %
the numbers of expected and observed events
validates the upstream background estimation procedure.

\section{Results}
\label{sec:Results}

\begin{figure}[tb]
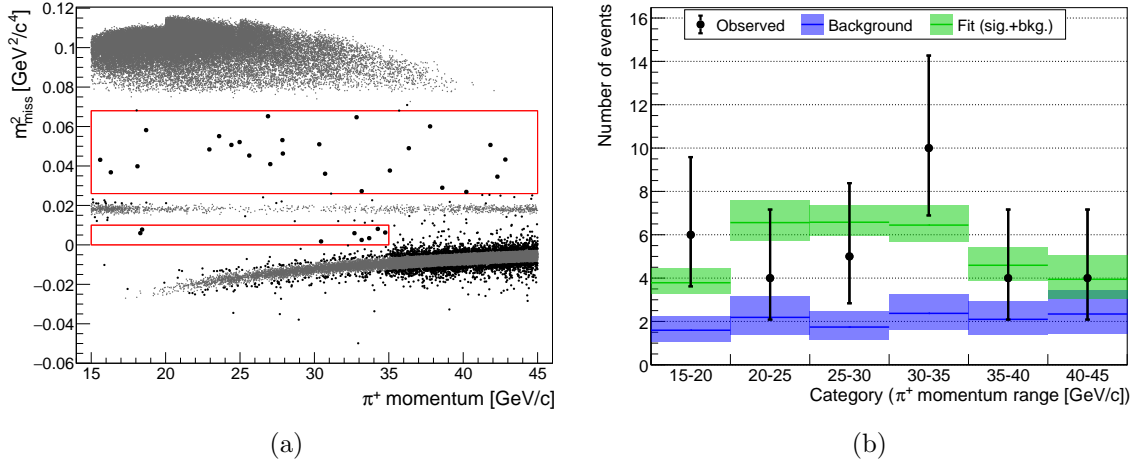

    \centering
    \begin{subfigure}{0.49\textwidth}
        \centering
        \includegraphics[width=\linewidth]{Figures/PNNUnmasking_finalTG_mm2_vs_p___Clean3_obs_all2D_BkgR_CRs_SRs___final_nogrid_Updated.pdf}
        \caption{}%
        \label{fig:Obs2324_mm2vsp}
    \end{subfigure}
    \hfill
    \begin{subfigure}{0.49\textwidth}
        \centering
        \includegraphics[width=\linewidth]{Figures/EvaluateResults_2324_postFIT__NinCategories_postFIT_PAPERfF_be_GRID_.pdf}
        \caption{}%
        \label{fig:Obs2324_Nfit}
    \end{subfigure}
    \caption{
        (a) Distribution in the $(p_\pi,m_\text{miss}^2)$ plane of the observed events in the 2023--2024 dataset satisfying the signal selection criteria.
        Events in the background, control and signal regions are shown by small grey,
        small black, and large black markers, respectively. Red boxes show the signal regions.
        (b) Numbers of expected and observed events in the six categories used for the fit of 2023--2024 data.%
    }
    \label{fig:Obs2324}
\end{figure}

After unmasking the signal regions, $33$ signal candidates are found in the 2023--2024 dataset, whereas $22.9\pm1.1$ SM signal events (table~\ref{tab:SESSummaryTable2324}) and $12.0^{+2.9}_{-2.2}$ background events (table~\ref{tab:BackgroundsSummaryTable2324}) are expected. 
The distribution of these events in the 
$(p_{\pi},m_{\rm miss}^{2})$ plane is shown in figure~\ref{fig:Obs2324_mm2vsp}.

The branching ratio is measured using a profile likelihood-ratio test statistic, $q$, taking into account the signal and background expectations with their uncertainties in each category. The uncertainties on the expectations are neglected for the determination of the statistical uncertainty. The systematic uncertainty is obtained from the total uncertainty by subtracting the statistical uncertainty in quadrature, and therefore is due to the uncertainties on the background and on the SES.
This approach is described in~\cite{Pnn2122paper}. 
For the 2023--2024 dataset, the six categories correspond to the six $p_\pi$ bins (section~\ref{sec:SignalSensitivity}), 
with expected and observed numbers of events in each category presented in table~\ref{tab:StatTreatmentInputTable2324}. 
The resulting measurement of the branching ratio is
\begin{equation}
\begin{aligned}
    \mathcal{B}_{2023-2024}(K^{+}\rightarrow\pi^{+}\nu\bar{\nu}) 
&= 
\left(
7.2
\left. {}^{+2.2}_{-1.9} \right|_{\text{stat}}
\left. {}^{+0.9}_{-0.9} \right|_{\text{syst}}
\right)
\times10^{-11}
\\&=
\left(
7.2^{+2.3}_{-2.1}\right)\times10^{-11} 
\,\,.
\end{aligned}
\label{eqn:BR2324}
\end{equation}
The comparison between the numbers of observed and expected events in each category is shown in figure~\ref{fig:Obs2324_Nfit}.
The goodness of fit is quantified by $q'_\text{min}/\text{ndf} = (-2\ln L'_\text{max}) / \text{ndf} = 4.42/5$, where $L'_\text{max}$ is the maximum likelihood in the hypothesis of independent Poisson-distributed observations (neglecting uncertainties on expectations), and $\text{ndf}$ is the number of degrees of freedom in the fit.

\begin{table}
\caption{
Numbers of observed ($N_\mathrm{obs}$) and expected SM signal ($N_{\pi\nu\nu}^\mathrm{SM}$) and background ($N_b$) events in each category for the 2023--2024 dataset.
The number of expected SM signal events is evaluated by equation~\eqref{eqn:Npnn} with $\mathcal{B}_{\pi\nu\bar{\nu}}^{\text{SM}} = 8.4\times10^{-11}$.
}
\vspace{8pt}
\centering
\begin{tabular}{|c|c||c||c|c|}
\hline 
Category ($\pi^+$ momentum range [$\text{GeV}/c$]) & $N_\text{obs}$ & $N_{\pi\nu\bar{\nu}}^{\text{SM}}$ & $N_{b}$ \T\B \\
\hline
\hline
 $15$--$20$ & $6$ & $2.57\pm0.12$ & $1.60^{+0.65}_{-0.51}$ \T\B\\
 $20$--$25$ & $4$ & $5.12\pm0.25$ & $2.19^{+1.00}_{-0.79}$ \T\B\\
 $25$--$30$ & $5$ & $5.66\pm0.27$ & $1.74^{+0.76}_{-0.57}$ \T\B\\
 $30$--$35$ & $10$ & $4.77\pm0.23$ & $2.37^{+0.91}_{-0.74}$ \T\B\\
 $35$--$40$ & $4$ & $2.92\pm0.14$ & $2.10^{+0.84}_{-0.69}$ \T\B\\
 $40$--$45$ & $4$ & $1.88\pm0.09$ & $2.34^{+1.10}_{-0.88}$ \T\B\\
\hline
\end{tabular}
\label{tab:StatTreatmentInputTable2324}
\end{table}

\begin{figure}[p]
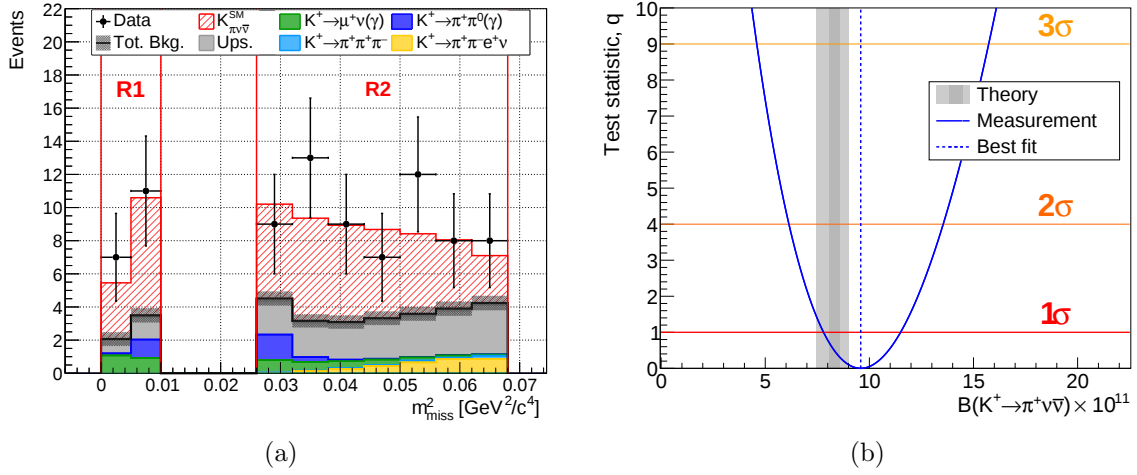

    \centering
    \begin{subfigure}{0.484\textwidth}
        \centering
        \includegraphics[width=\linewidth]{Figures/PnnSpectrumStudy__1624_mm2_SB_HStack_FINAL2_vF.pdf}
        \caption{}%
        \label{fig:Obs1624_mm2}
    \end{subfigure}
    \hfill
    \begin{subfigure}{0.49\textwidth}
        \centering
        \includegraphics[width=\linewidth]{Figures/EvaluateResults_NA62_FINAL_NEW_dq_vs_Br_FullOnly__SM_TheoryEnvolope_F2aR_PaperFINAL.pdf} %
        \caption{}%
        \label{fig:Obs1624_BR_Likelihood}
    \end{subfigure}
    \caption{
        (a) Distribution in $m_\text{miss}^2$ of the observed events in the 2016--2024 dataset, with SM signal expectation and estimated backgrounds. The total background and its uncertainty are shown by the black line and hatched area, respectively. 
        R1 and R2 denote the two signal regions.
        (b) Test statistic as a function of the $K^+\to\pi^+\nu\bar\nu$ branching ratio for 2016--2024 data and recent SM predictions~\cite{Buras:2022wpw,KaonsAtCERN2023,Allwicher:2024ncl}. The dark grey band shows the range of the predictions, while the light grey band shows the range of their uncertainties.
    }
    \label{fig:Obs1624}
\end{figure}

The six categories from 2023--2024 data are combined with the fifteen resulting from the previous analyses of the 2016--2022 data~\cite{Pnn2122paper}. For the 2016--2024 dataset, with an expectation of $30^{+4}_{-3}$ background events and $43\pm1$ SM signal events, an expected significance for the SM observation exceeding $5\sigma$ is reached for the first time. 
Eighty-four events are observed in the 2016--2024 dataset,
leading to a significance for the rejection of the background-only hypothesis exceeding $6\sigma$. 
The $m_\text{miss}^2$ distribution for the observed events, together with those for the expectations of SM signal and background, is shown in figure~\ref{fig:Obs1624_mm2}.
The fit procedure is repeated using the $21$ categories spanning the 2016--2024 dataset.
The test statistic is shown in figure~\ref{fig:Obs1624_BR_Likelihood}, as a function of the branching ratio, and the measurement is
\begin{equation}
\begin{aligned}
    \mathcal{B}_{2016-2024}(K^{+}\rightarrow\pi^{+}\nu\bar{\nu}) 
&=
\left(
9.6
\left.{}^{+ 1.8}_{ - 1.6} \right|_{\text{stat}} \left.{}^{+ 0.8}_{- 0.6} \right|_{\text{syst}}  \right)\times10^{-11}
\\
&= 
\left(9.6^{+ 1.9}_{- 1.8} \right)\times10^{-11} 
\,\,,
\end{aligned}
\label{eqn:BR1624}
\end{equation}
with a goodness of fit $q'_\text{min}/\text{ndf}=15.2/20$.

\begin{figure}[p]
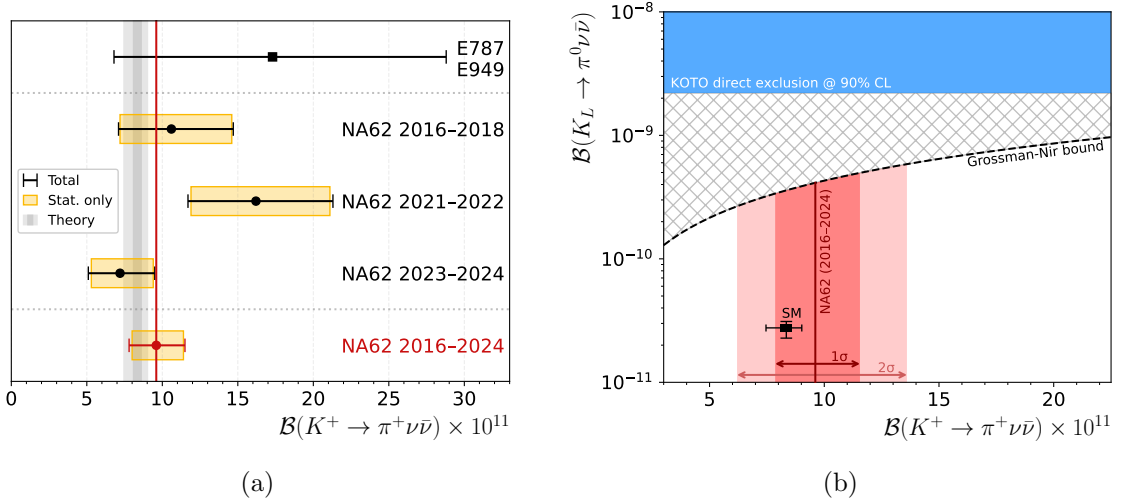

    \centering
    \begin{subfigure}{0.49\textwidth}
        \centering
        \includegraphics[width=0.935\linewidth]{Figures/br_history_v2.pdf}\hfill
        \caption{}%
        \label{fig:ResultsInContext_1D}
    \end{subfigure}
    \hfill
    \begin{subfigure}{0.49\textwidth}
        \centering
        \includegraphics[width=\linewidth]{Figures/PlotPNNBrResults_BrKL_vs_BrKP_logy_Envolope_Grid0_annotated_errorbars.pdf}
        \caption{}%
        \label{fig:ResultsInContext_2D}
    \end{subfigure}
    \caption{
        (a) Summary of $K^{+}\rightarrow\pi^{+}\nu\bar{\nu}$ branching ratio measurements, compared to 
         SM predictions~\cite{Buras:2022wpw,KaonsAtCERN2023,Allwicher:2024ncl}.
         The dark grey band shows the range of the predictions, while the light grey band shows the range of their uncertainties.
        (b) Current constraints on $K\rightarrow\pi\nu\bar{\nu}$ decay modes. The black rectangular marker covers the range of recent SM predictions~\cite{Buras:2022wpw,KaonsAtCERN2023,Allwicher:2024ncl}, while the bars bracket their uncertainties.
        The $1\sigma$ and $2\sigma$ confidence intervals for the NA62 result are displayed in the darker and lighter shaded red areas, respectively.
    }
    \label{fig:ResultsInContext}
\end{figure}

Figure~\ref{fig:ResultsInContext_1D} summarizes the $K^{+}\rightarrow\pi^{+}\nu\bar{\nu}$ branching ratio measurements from the BNL experiments E787 and E949~\cite{BNL-E949:2009dza}, and from the NA62 experiment with data from 2016--2018~\cite{PnnRun1Paper}, 2021--2022~\cite{Pnn2122paper} and 
2023--2024, together with the combined NA62 result. %
The current constraints on the $K\to\pi\nu\bar\nu$ decay modes, including the most stringent direct upper limit on $\mathcal{B}(K_{L}\rightarrow\pi^{0}\nu\bar{\nu})$~\cite{KOTO:2024zbl} and the theoretical model-independent Grossman-Nir bound~\cite{Grossman:1997sk,PDG}, are illustrated in figure~\ref{fig:ResultsInContext_2D}.

\section{Conclusions}

The measurement $\mathcal{B}(K^{+}\rightarrow\pi^{+}\nu\bar{\nu}) = \left(9.6^{+ 1.9}_{- 1.8} \right)\times10^{-11}$, obtained with the 2016--2024 dataset, is compatible with SM predictions and achieves a relative uncertainty of $20\%$.
The inclusion of the dataset collected in 2023--2024 under optimised conditions provides a major improvement in precision over the measurement using 2016--2022 data, doubling the effective sample size while significantly decreasing the relative background contribution.

The NA62 experiment collected data in 2025 and will conclude data taking in 2026. 
Based on the number of days of data taking, the number of signal events in the final dataset is expected to increase by a further $50\%$.

\clearpage
\section*{Acknowledgements}
It is a pleasure to express our appreciation to the staff of the CERN laboratory and the technical
staff of the participating laboratories and universities for their efforts in the operation of the
experiment and data processing.

The cost of the experiment and its auxiliary systems was supported by the funding agencies of 
the Collaboration Institutes. We are particularly indebted to: 
F.R.S.-FNRS (Fonds de la Recherche Scientifique - FNRS), under Grants No. 4.4512.10, 1.B.258.20, Belgium;
CECI (Consortium des Equipements de Calcul Intensif), funded by the Fonds de la Recherche Scientifique de Belgique (F.R.S.-FNRS) under Grant No. 2.5020.11 and by the Walloon Region, Belgium;
NSERC (Natural Sciences and Engineering Research Council), funding SAPPJ-2020-00026 and SAPPJ-2024-00027, Canada;
MEYS (Ministry of Education, Youth and Sports) funding LM 2018104, Czech Republic;
BMFTR (Bundesministerium f\"ur Forschung, Technologie und Raumfahrt), grant No. 05H24UM5, Germany;
INFN  (Istituto Nazionale di Fisica Nucleare),  Italy;
MIUR (Ministero dell'Istruzione, dell'Universit\`a e della Ricerca),  Italy;
CONACyT  (Consejo Nacional de Ciencia y Tecnolog\'{i}a),  Mexico;
IFA (Institute of Atomic Physics) Romanian 
CERN-RO Nr.\break 06/03.01.2022
and Nucleus Programme PN 19 06 01 04,  Romania;
MESRS  (Ministry of Education, Science, Research and Sport), Slovakia; 
CERN (European Organization for Nuclear Research), Switzerland; 
SNSF (Swiss National Science Foundation), grant No. 219639, Switzerland;
STFC (Science and Technology Facilities Council), United Kingdom;
NSF (National Science Foundation) Award Numbers 1506088 and 1806430,  U.S.A.;
ERC (European Research Council)  ``UniversaLepto'' advanced grant 268062, ``KaonLepton'' starting grant 336581, Europe.

Individuals have received support from:
Charles University (grants UNCE 24/SCI/016, PRIMUS 23/SCI/025), 
Ministry of Education, Youth and Sports (project FORTE\break
CZ.02.01.01/00/22-008/0004632), Czech Republic;
Czech Science Foundation (grant 23-06770S);   
Agence Nationale de la Recherche (grant ANR-19-CE31-0009), France;
Deutsche Forschungsgemeinschaft (DFG Excellence Strategy  - EXC 2094 - 390783311), Germany;
Ministero dell'Istruzione, dell'Universit\`a e della Ricerca (MIUR  ``Futuro in ricerca 2012''  grant RBFR12JF2Z, Project GAP), Italy;
Nuclemedica Soluciones, San Luis Potos\'{i}, Mexico;
the Royal Society  (grants UF100308, UF0758946), United Kingdom;
STFC (Rutherford fellowships ST/J00412X/1, ST/M005798/1), United Kingdom;
ERC (grants 268062,  336581 and  starting grant 802836 ``AxScale'');
EU Horizon 2020 (Marie Sk\l{}odowska-Curie grants 701386, 754496, 842407, 893101, 101023808).

\newpage
\clearpage
\bibliography{bibliography}
\newpage
\clearpage

\newcommand{\orcimg}{\raisebox{-0.3\height}{\includegraphics[height=\fontcharht\font`A]{ORCIDiD_iconvector}}}
\newcommand{\orcid}[1]{\href{https://orcid.org/#1}{\orcimg}}

\addcontentsline{toc}{section}{The NA62 Collaboration}
\centerline{\bf The NA62 Collaboration}
\vspace{0.5cm}

\begin{raggedright}
\noindent
{\bf UCLouvain, Centre for Cosmology, Particle Physics and Phenomenology, CP3, Louvain-La-Neuve, Belgium}\\
 S.~Alibocus\orcid{0000-0002-9377-8852},
 B.~Bloch-Devaux$\,${\footnotemark[1]}\orcid{0000-0002-2463-1232},
 E.~Cortina Gil\orcid{0000-0001-9627-699X},
 N.~Lurkin\orcid{0000-0002-9440-5927}
\vspace{0.5cm}

{\bf TRIUMF, Vancouver, British Columbia, Canada}\\
 T.~Numao\orcid{0000-0001-5232-6190},
 V.~Shang\orcid{0000-0002-1436-6092},
 B.~Velghe\orcid{0000-0002-0797-8381},
 V. W. S.~Wong\orcid{0000-0001-5975-8164}
\vspace{0.5cm}

{\bf University of British Columbia, Vancouver, British Columbia, Canada}\\
 D.~Bryman$\,${\footnotemark[2]}\orcid{0000-0002-9691-0775}
\vspace{0.5cm}

{\bf Charles University, Prague, Czech Republic}\\
 L.~Bician\orcid{0000-0001-9318-0116},
 M.~Cirkovic\orcid{0000-0002-4420-9688},
 Z.~Hives$\,${\footnotemark[3]}\orcid{0000-0002-5025-993X},
 T.~Husek$\,${\footnotemark[1]}\orcid{0000-0002-7208-9150},
 K.~Kampf\orcid{0000-0003-1096-667X}, 
 M.~Kolesar\orcid{0000-0002-9085-2252},
 M.~Koval\orcid{0000-0002-6027-317X},
 M.~Lelak\orcid{0009-0006-1788-1993},
 E.~Long\orcid{0000-0002-7192-2235}
\vspace{0.5cm}

{\bf Aix Marseille University, CNRS/IN2P3, CPPM, Marseille, France}\\
 B.~De Martino\orcid{0000-0003-2028-9326},
 M.~Perrin-Terrin\orcid{0000-0002-3568-1956},
 L.~Petit$\,${\footnotemark[4]}\orcid{0009-0000-8079-9710}
\vspace{0.5cm}

{\bf Max-Planck-Institut f\"ur Physik (Werner-Heisenberg-Institut), Garching, Germany}\\
 B.~D\"obrich\orcid{0000-0002-6008-8601},
 D.~Grewe\orcid{0009-0008-0953-8436},
 J.~Jerhot\orcid{0000-0002-3236-1471},
 S.~Lezki\orcid{0000-0002-6909-774X},
 J.~Schubert$\,${\footnotemark[5]}\orcid{0000-0002-5782-8816}
\vspace{0.5cm}

{\bf Institut f\"ur Physik and {\boldmath{PRISMA$^+$} Cluster of Excellence, Universit\"at Mainz, Mainz, Germany}}\\
 A. T.~Akmete\orcid{0000-0002-5580-5477},
 R.~Aliberti$\,${\footnotemark[6]}\orcid{0000-0003-3500-4012},
 M.~Ceoletta$\,${\footnotemark[7]}\orcid{0000-0002-2532-0217},
 L.~Di Lella\orcid{0000-0003-3697-1098},
 N.~Doble\orcid{0000-0002-0174-5608}, 
 L.~Peruzzo\orcid{0000-0002-4752-6160},
 C.~Polivka\orcid{0009-0002-2403-8575},
 S.~Schuchmann\orcid{0000-0002-8088-4226},
 H.~Wahl\orcid{0000-0003-0354-2465},
 R.~Wanke\orcid{0000-0002-3636-360X}
\vspace{0.5cm}

{\bf INFN, Sezione di Ferrara, Ferrara, Italy}\\
 L.~Bandiera\orcid{0000-0002-5537-9674},
 A.~Gianoli\orcid{0000-0002-2456-8667}
\vspace{0.5cm}

{\bf INFN, Sezione di Ferrara e Dipartimento di Fisica e Scienze della Terra dell'Universit\`a, Ferrara, Italy}\\
 N.~Canale\orcid{0000-0003-2262-7077},
 P.~Fedeli\orcid{0009-0002-0570-6048},
 R.~Negrello\orcid{0009-0008-3396-5550},
 I.~Neri\orcid{0000-0002-9669-1058},
 F.~Petrucci\orcid{0000-0002-7220-6919},
 M.~Romagnoni\orcid{0000-0002-2775-6903}
\vspace{0.5cm}

{\bf INFN, Sezione di Firenze, Sesto Fiorentino, Italy}\\
 A.~Bizzeti$\,${\footnotemark[8]}\orcid{0000-0001-5729-5530},
 F.~Bucci\orcid{0000-0003-1726-3838}
\vspace{0.5cm}

{\bf INFN, Sezione di Firenze e Dipartimento di Fisica e Astronomia dell'Universit\`a, Sesto Fiorentino, Italy}\\
 M.~Lenti\orcid{0000-0002-2765-3955},
 P.~Lo Chiatto$\,${\footnotemark[9]}\orcid{0000-0002-4177-557X},
 I.~Panichi\orcid{0000-0001-7749-7914},
 G.~Ruggiero\orcid{0000-0001-6605-4739}
\vspace{0.5cm}

{\bf INFN, Laboratori Nazionali di Frascati, Frascati, Italy}\\
 A.~Antonelli\orcid{0000-0001-7671-7890},
 V.~Kozhuharov$\,${\footnotemark[10]}\orcid{0000-0002-0669-7799},
 G.~Lanfranchi\orcid{0000-0002-9467-8001},
 S.~Martellotti\orcid{0000-0002-4363-7816},
 M.~Moulson\orcid{0000-0002-3951-4389}, 
 L.~Plini$\,${\footnotemark[11]}\orcid{0009-0004-0498-1333},
 M.~Soldani$\,${\footnotemark[3]}\orcid{0000-0003-4902-943X},
 T.~Spadaro\orcid{0000-0002-7101-2389},
 J.~Swallow$\,$\renewcommand{\thefootnote}{\fnsymbol{footnote}}\footnotemark[1]\renewcommand{\thefootnote}{\arabic{footnote}}$^,$$\,${\footnotemark[3]}\orcid{0000-0002-1521-0911},
 G.~Tinti\orcid{0000-0003-1364-844X}
\vspace{0.5cm}

{\bf INFN, Sezione di Napoli e Dipartimento di Fisica ``Ettore Pancini'', Napoli, Italy}\\
 F.~Ambrosino\orcid{0000-0001-5577-1820},
 M.~D'Errico\orcid{0000-0001-5326-1106},
 R.~Fiorenza$\,$\renewcommand{\thefootnote}{\fnsymbol{footnote}}\footnotemark[1]\renewcommand{\thefootnote}{\arabic{footnote}}$^,$$\,${\footnotemark[3]}\orcid{0000-0003-4965-7073},
 M.~Francesconi\orcid{0000-0002-7029-7634},
 R.~Giordano\orcid{0000-0002-5496-7247}, 
 P.~Massarotti\orcid{0000-0002-9335-9690},
 M.~Mirra\orcid{0000-0002-1190-2961},
 M.~Napolitano\orcid{0000-0003-1074-9552},
 I.~Rosa$\,${\footnotemark[12]}\orcid{0009-0002-7564-1825},
 G.~Saracino\orcid{0000-0002-0714-5777}
\vspace{0.5cm}

{\bf INFN, Sezione di Perugia, Perugia, Italy}\\
 P.~Cenci\orcid{0000-0001-6149-2676},
 V.~Duk\orcid{0000-0001-6440-0087},
 V.~Falaleev\orcid{0000-0003-3150-2196},
 R.~Lollini\orcid{0000-0003-3898-7464},
 P.~Lubrano\orcid{0000-0003-0221-4806}, 
 M.~Pepe\orcid{0000-0001-5624-4010},
 M.~Piccini\orcid{0000-0001-8659-4409}
\vspace{0.5cm}

{\bf INFN, Sezione di Perugia e Dipartimento di Fisica e Geologia dell'Universit\`a, Perugia, Italy}\\
 G.~Anzivino\orcid{0000-0002-5967-0952}
\vspace{0.5cm}

{\bf INFN, Sezione di Pisa, Pisa, Italy}\\
 R.~Fantechi\orcid{0000-0002-6243-5726},
 S.~Kholodenko$\,${\footnotemark[3]}\orcid{0000-0002-0260-6570},
 F.~Spinella\orcid{0000-0002-9607-7920}
\vspace{0.5cm}

{\bf INFN, Sezione di Pisa e Dipartimento di Fisica dell'Universit\`a, Pisa, Italy}\\
 F.~Costantini\orcid{0000-0002-2974-0067},
 M.~Giorgi\orcid{0000-0001-9571-6260},
 S.~Giudici\orcid{0000-0003-3423-7981},
 G.~Lamanna\orcid{0000-0001-7452-8498},
 E.~Lari\orcid{0000-0003-3303-0524}, 
 E.~Pedreschi\orcid{0000-0001-7631-3933},
 J.~Pinzino\orcid{0000-0002-7418-0636},
 M.~Sozzi\orcid{0000-0002-2923-1465}
\vspace{0.5cm}

{\bf INFN, Sezione di Pisa e Scuola Normale Superiore, Pisa, Italy}\\
 I.~Mannelli\orcid{0000-0003-0445-7422}
\vspace{0.5cm}

{\bf INFN, Sezione di Roma I, Roma, Italy}\\
 A.~Biagioni\orcid{0000-0001-5820-1209},
 P.~Cretaro\orcid{0000-0002-2229-149X},
 O.~Frezza\orcid{0000-0001-8277-1877},
 F.~Lo Cicero\orcid{0000-0002-9904-2619},
 A.~Lonardo\orcid{0000-0002-5909-6508}, 
 L.~Pontisso\orcid{0000-0001-7137-5254},
 F.~Simula\orcid{0000-0002-7955-1491},
 M.~Turisini\orcid{0000-0002-5422-1891},
 P.~Vicini\orcid{0000-0002-4379-4563}
\vspace{0.5cm}

{\bf INFN, Sezione di Roma I e Dipartimento di Fisica, Sapienza Universit\`a di Roma, Roma, Italy}\\
 C.~Chiarini\orcid{0009-0008-1278-7708},
 M.~Raggi\orcid{0000-0002-7448-9481},
 C.~Rossi\orcid{0000-0001-5716-1401}
\vspace{0.5cm}

{\bf INFN, Sezione di Roma Tor Vergata, Roma, Italy}\\
 R.~Ammendola\orcid{0000-0003-4501-3289},
 V.~Bonaiuto$\,${\footnotemark[13]}\orcid{0000-0002-2328-4793},
 A.~Fucci,
 A.~Salamon\orcid{0000-0002-8438-8983},
 F.~Sargeni$\,${\footnotemark[14]}\orcid{0000-0002-0131-236X}
\vspace{0.5cm}

{\bf INFN, Sezione di Torino, Torino, Italy}\\
 C.~Biino$\,${\footnotemark[15]}\orcid{0000-0002-1397-7246},
 A.~Filippi\orcid{0000-0003-4715-8748},
 F.~Marchetto\orcid{0000-0002-5623-8494},
 D.~Soldi\orcid{0000-0001-9059-4831}
\vspace{0.5cm}

{\bf INFN, Sezione di Torino e Dipartimento di Fisica dell'Universit\`a, Torino, Italy}\\
 R.~Arcidiacono$\,${\footnotemark[16]}\orcid{0000-0001-5904-142X},
 E.~Menichetti\orcid{0000-0001-7143-8200},
 E.~Migliore\orcid{0000-0002-2271-5192}
\vspace{0.5cm}

{\bf Institute of Nuclear Physics, Almaty, Kazakhstan}\\
 Y.~Mukhamejanov\orcid{0000-0002-9064-6061},
 A.~Mukhamejanova$\,${\footnotemark[17]}\orcid{0009-0004-4799-9066},
 N.~Saduyev\orcid{0000-0002-5144-0677},
 S.~Sakhiyev\orcid{0000-0002-9014-9487}
\vspace{0.5cm}

\newpage
{\bf Instituto de F\'isica, Universidad Aut\'onoma de San Luis Potos\'i, San Luis Potos\'i, Mexico}\\
 A.~Briano Olvera\orcid{0000-0001-6121-3905},
 A. E.~D\'{\i}az Rodarte$\,${\footnotemark[9]}\orcid{0009-0006-4038-4792},
 J.~Engelfried\orcid{0000-0001-5478-0602},
 N.~Estrada-Tristan$\,${\footnotemark[18]}\orcid{0000-0003-2977-9380},
 A.~E.~Mart\'{\i}nez Hern\'andez\orcid{0009-0005-5180-8039}, 
 R.~Piandani\orcid{0000-0003-2226-8924},
 M.~A.~Reyes Santos$\,${\footnotemark[18]}\orcid{0000-0003-1347-2579},
 C.~A.~Rico Olvera$\,${\footnotemark[18]}\orcid{0009-0008-0451-9439},
 K.~A.~Rodriguez Rivera\orcid{0000-0001-5723-9176}
\vspace{0.5cm}

{\bf Horia Hulubei National Institute for R\&D in Physics and Nuclear Engineering, Bucharest-Magurele, Romania}\\
 P.~Boboc\orcid{0000-0001-5532-4887},
 A.~M.~Bragadireanu,
 S.~A.~Ghinescu\orcid{0000-0003-3716-9857},
 O.~E.~Hutanu
\vspace{0.5cm}

{\bf Faculty of Mathematics, Physics and Informatics, Comenius University, Bratislava, Slovakia}\\
 T.~Blazek\orcid{0000-0002-2645-0283},
 V.~Cerny\orcid{0000-0003-1998-3441},
 T.~Velas\orcid{0009-0004-0061-1968},
 R.~Volpe$\,${\footnotemark[19]}\orcid{0000-0003-1782-2978}
\vspace{0.5cm}

{\bf CERN, European Organization for Nuclear Research, Geneva, Switzerland}\\
 J.~Bernhard\orcid{0000-0001-9256-971X},
 M.~Boretto\orcid{0000-0001-5012-4480},
 F.~Brizioli$\,${\footnotemark[20]}\orcid{0000-0002-2047-441X},
 A.~Ceccucci\orcid{0000-0002-9506-866X},
 M.~Corvino\orcid{0000-0002-2401-412X}, 
 H.~Danielsson\orcid{0000-0002-1016-5576},
 F.~Duval,
 L.~Federici$\,${\footnotemark[21]}\orcid{0000-0002-3401-9522},
 E.~Gamberini\orcid{0000-0002-6040-4985},
 R.~Guida\orcid{0000-0001-8413-9672}, 
 E. B.~Holzer\orcid{0000-0003-2622-6844},
 B.~Jenninger,
 Z.~Kucerova$\,${\footnotemark[22]}\orcid{0000-0001-8906-3902},
 G.~Lehmann Miotto\orcid{0000-0001-9045-7853},
 P.~Lichard\orcid{0000-0003-2223-9373}, 
 E.~Minucci$\,${\footnotemark[23]}$^,$$\,${\footnotemark[24]}\orcid{0000-0002-3972-6824},
 M.~Noy,
 V.~Ryjov,
 M.~Zamkovsky$\,${\footnotemark[25]}\orcid{0000-0002-5067-4789}
\vspace{0.5cm}

{\bf Ecole Polytechnique F\'ed\'erale Lausanne, Lausanne, Switzerland}\\
 G.~Belleri\orcid{0009-0006-4472-8142},
 X.~Chang$\,$\renewcommand{\thefootnote}{\fnsymbol{footnote}}\footnotemark[1]\renewcommand{\thefootnote}{\arabic{footnote}}\orcid{0000-0002-8792-928X},
 L. M.~Garc\'{\i}a Mart\'{\i}n\orcid{0000-0003-0714-8991},
 A.~Kleimenova\orcid{0000-0002-9129-4985},
 R.~Marchevski\orcid{0000-0003-3410-0918}
\vspace{0.5cm}

{\bf School of Physics and Astronomy, University of Birmingham, Birmingham, United Kingdom}\\
 J. R.~Fry\orcid{0000-0002-3680-361X},
 F.~Gonnella\orcid{0000-0003-0885-1654},
 E.~Goudzovski\orcid{0000-0001-9398-4237},
 J.~Henshaw\orcid{0000-0001-7059-421X},
 C.~Kenworthy\orcid{0009-0002-8815-0048}, 
 C.~Lazzeroni\orcid{0000-0003-4074-4787},
 C.~Parkinson\orcid{0000-0003-0344-7361},
 A.~Romano$\,${\footnotemark[26]}\orcid{0000-0003-1779-9122},
 C.~Sam\orcid{0009-0005-3802-5777},
 J.~Sanders\orcid{0000-0003-1014-094X}, 
 A.~Sergi$\,${\footnotemark[27]}\orcid{0000-0001-9495-6115},
 A.~Tomczak\orcid{0000-0001-5635-3567}
\vspace{0.5cm}

{\bf School of Physics, University of Bristol, Bristol, United Kingdom}\\
 H.~Heath\orcid{0000-0001-6576-9740}
\vspace{0.5cm}

{\bf School of Physics and Astronomy, University of Glasgow, Glasgow, United Kingdom}\\
 D.~Britton\orcid{0000-0001-9998-4342},
 A.~Norton\orcid{0000-0001-5959-5879},
 D.~Protopopescu\orcid{0000-0002-8047-6513}
\vspace{0.5cm}

{\bf School of Physics and Astronomy, University of Lancaster, Lancaster, United Kingdom}\\
 J.~Cook\orcid{0009-0004-9930-9055},
 J. B.~Dainton\orcid{0009-0002-5745-5375},
 M. D.~Galati\orcid{0000-0002-8716-4440},
 L.~Gatignon\orcid{0000-0001-6439-2945},
 R. W. L.~Jones\orcid{0000-0002-6427-3513}, 
 K.~Massri\orcid{0000-0001-7533-6295},
 A.~Shaikhiev\orcid{0000-0003-2921-8743},
 M.~Thompson-Walker\orcid{0009-0002-2374-4769}
\vspace{0.5cm}

{\bf Physics and Astronomy Department, George Mason University, Fairfax, Virginia, USA}\\
 P.~Cooper,
 D.~Coward$\,${\footnotemark[28]}\orcid{0000-0001-7588-1779},
 P.~Rubin\orcid{0000-0001-6678-4985}
\vspace{0.5cm}

\newpage
{\bf Authors affiliated with an international laboratory covered by a cooperation agreement with CERN}\\
 A.~Baeva,
 D.~Baigarashev$\,${\footnotemark[29]}\orcid{0000-0001-6101-317X},
 V.~Bautin\orcid{0000-0002-5283-6059},
 D.~Emelyanov,
 T.~Enik\orcid{0000-0002-2761-9730}, 
 V.~Kekelidze\orcid{0000-0001-8122-5065},
 D.~Kereibay,
 A.~Korotkova,
 L.~Litov$\,${\footnotemark[10]}\orcid{0000-0002-8511-6883},
 D.~Madigozhin\orcid{0000-0001-8524-3455}, 
 M.~Misheva$\,${\footnotemark[30]},
 N.~Molokanova,
 I.~Polenkevich,
 Yu.~Potrebenikov\orcid{0000-0003-1437-4129},
 K.~Salamatin\orcid{0000-0001-6287-8685}, 
 S.~Shkarovskiy
\vspace{0.5cm}

{\bf Authors affiliated with an Institute formerly covered by a cooperation agreement with CERN}\\
 S.~Fedotov,
 K.~Gorshanov\orcid{0000-0001-7912-5962},
 E.~Gushchin\orcid{0000-0001-8857-1665},
 A.~Khotyantsev,
 Y.~Kudenko\orcid{0000-0003-3204-9426}, 
 V.~Kurochka,
 V.~Kurshetsov\orcid{0000-0003-0174-7336},
 A.~Mefodev,
 V.~Obraztsov\orcid{0000-0002-0994-3641},
 A.~Okhotnikov\orcid{0000-0003-1404-3522}, 
 A.~Sadovskiy\orcid{0000-0002-4448-6845},
 V.~Sugonyaev\orcid{0000-0003-4449-9993},
 O.~Yushchenko\orcid{0000-0003-4236-5115}
\vspace{0.5cm}

\end{raggedright}

\setcounter{footnote}{0}
\newlength{\basefootnotesep}
\setlength{\basefootnotesep}{\footnotesep}

\renewcommand{\thefootnote}{\fnsymbol{footnote}}
\noindent
$^{\footnotemark[1]}${Corresponding authors: X.~Chang, R.~Fiorenza, J.~Swallow, \\
 email:  xiafei.chang@cern.ch, renato.fiorenza@cern.ch, joel.christopher.swallow@cern.ch}\\
\renewcommand{\thefootnote}{\arabic{footnote}}
$^{1}${Also at School of Physics and Astronomy, University of Birmingham, Birmingham, B15 2TT, UK} \\
$^{2}${Also at TRIUMF, Vancouver, British Columbia, V6T 2A3, Canada} \\
$^{3}${Present address: CERN, European Organization for Nuclear Research, CH-1211 Geneva 23, Switzerland} \\
$^{4}${Also at Universit\'e de Toulon, Aix Marseille University, CNRS, IM2NP, F-83957, La Garde, France} \\
$^{5}${Also at Department of Physics, Technical University of Munich, M\"unchen, D-80333, Germany} \\
$^{6}${Present address: Institut f\"ur Kernphysik and Helmholtz Institute Mainz, Universit\"at Mainz, Mainz, D-55099, Germany} \\
$^{7}${Also at CERN, European Organization for Nuclear Research, CH-1211 Geneva 23, Switzerland} \\
$^{8}${Also at Dipartimento di Scienze Fisiche, Informatiche e Matematiche, Universit\`a di Modena e Reggio Emilia, I-41125 Modena, Italy} \\
$^{9}${Present address: Max-Planck-Institut f\"ur Physik (Werner-Heisenberg-Institut), Garching, D-85748, Germany} \\
$^{10}${Also at Faculty of Physics, University of Sofia, BG-1164 Sofia, Bulgaria} \\
$^{11}${Also at Dipartimento di Ingegneria Informatica Automatica e Gestionale, Universit\`a di Roma La Sapienza, I-00185 Roma, Italy} \\
$^{12}${Present address: INFN, Sezione di Napoli e Scuola Superiore Meridionale, I-80138 Napoli, Italy} \\
$^{13}${Also at Department of Industrial Engineering, University of Roma Tor Vergata, I-00173 Roma, Italy} \\
$^{14}${Also at Department of Electronic Engineering, University of Roma Tor Vergata, I-00173 Roma, Italy} \\
$^{15}${Also at Gran Sasso Science Institute, I-67100 L'Aquila,  Italy} \\
$^{16}${Also at Universit\`a degli Studi del Piemonte Orientale, I-13100 Vercelli, Italy} \\
$^{17}${Also at al-Farabi Kazakh National University, 050040 Almaty, Kazakhstan} \\
$^{18}${Also at Universidad de Guanajuato, 36000 Guanajuato, Mexico} \\
$^{19}${Present address: Charles University, 116 36 Prague 1, Czech Republic} \\
$^{20}${Present address: INFN, Sezione di Perugia e Dipartimento di Fisica e Geologia\break dell'Universit\`a, I-06100 Perugia, Italy} \\
$^{21}${Present address: IPHC, CNRS/IN2P3, Strasbourg University, F-67037 Strasbourg,\break France} \\
$^{22}${Present address: Faculty of Mathematics, Physics and Informatics, Comenius University, 842 48, Bratislava, Slovakia} \\
$^{23}${Present address: INFN, Laboratori Nazionali di Frascati, I-00044 Frascati, Italy} \\
$^{24}${Also at Museo Storico della Fisica e Centro Studi e Ricerche Enrico Fermi, I-00184  Roma, Italy} \\
$^{25}${Present address: UCLouvain, Centre for Cosmology, Particle Physics and Phenomenology, CP3, B-1348 Louvain-La-Neuve, Belgium} \\
$^{26}${Present address: Department of Physics, University of Warwick, Coventry, CV4 7AL, UK} \\
$^{27}${Present address: INFN, Sezione di Genova e Dipartimento di Fisica dell'Universit\`a, I-16146 Genova, Italy} \\
$^{28}${Also at SLAC National Accelerator Laboratory, Stanford University, Menlo Park, CA 94025, USA} \\
$^{29}${Also at L. N. Gumilyov Eurasian National University, 010000 Astana, Kazakhstan} \\
$^{30}${Present address: Institute of Nuclear Research and Nuclear Energy of Bulgarian Academy of Science (INRNE-BAS), BG-1784 Sofia, Bulgaria} \\

\immediate\closein\imgstream

\end{document}